\RequirePackage{fix-cm}
\documentclass{svjour3}                     
\smartqed  
\usepackage{graphicx}
\usepackage{bm}
\usepackage{amsmath}
\usepackage{amssymb}
\usepackage{mathptmx}      
\usepackage{latexsym}
\usepackage{verbatim}
\usepackage{color}
\usepackage[english]{babel}

\journalname{Journal of Computational Electronics}
\begin{document}
\title{Quantum transport of pseudospin-polarized Dirac fermions in gapped graphene nanostructures}

\author{Leyla Majidi \and Malek Zareyan
}


\institute{Leyla Majidi  \and
           Malek Zareyan  \at
           Department of Physics, Institute for Advanced Studies in Basic Sciences (IASBS), P. O. Box 45195-1159, Zanjan 45137-66731, Iran\\
           \email{zareyan@iasbs.ac.ir}
}

\date{Received: date / Accepted: date}

\maketitle

\begin{abstract}
We investigate the unusual features of the quantum transport in gapped monolayer graphene, which is in a pseudospin symmetry-broken state with a net perpendicular pseudomagnetization. Using these pseudoferromagnets (PFs), we propose a perfect pseudospin valve effect that can be used for realizing pseudospintronics in monolayer graphene. The peculiarity of the associated effects of pseudo spin injection and pseudo spin accumulation are also studied. We further demonstrate the determining effect of the sublattice pseudospin degree of freedom on Andreev reflection and the associated proximity effect in hybrid structures of PFs and a superconductor in S/PF and PF/S/PF geometries. In particular, we find a peculiar Andreev reflection that is associated with an inversion of the $z$ component of the carriers pseudospin vector. Our results show that the gapped normal graphene behaves like a ferromagnetic graphene and the effect of the pseudospin degree of freedom in gapped graphene is as important as the spin in a ferromagnetic graphene.

\keywords{Sublattice pseudospin \and Gapped graphene \and Pseudospin valve \and Superconducting proximity effect \and Andreev reflection}
\PACS{72.80.Vp \and 72.25.-b \and 85.75.-d \and 74.45.+c}
\end{abstract}

\section{\label{sec:intro}Introduction}
Graphene, the two dimensional layer of the carbon atoms with honeycomb lattice structure, has attracted a great deal of attention as
a new promising material for nanoelectronics, since its experimental realization a few years ago\cite{Novoselov04,Novoselov05,Zhang05}. Graphene has a zero-gap semiconducting band structure in which the charge carriers behave like 2D massless Dirac fermions with a pseudo-relativistic chiral property. The carrier type, [electron-like ($n$) or hole-like ($p$)] and its density can be tuned by means of electrical gate or doping of underlying substrate. Most of the peculiar properties of graphene is the result of its massless Dirac spectrum of the low-lying electron-hole excitations, which
in addition to the regular spin appear to come endowed with the two quantum degrees of freedom, the so called pseudospin and valley. The
pseudospin represents the sublattice degree of freedom of the graphene's honeycomb structure, and the valley defines the corresponding
degree of freedom in the reciprocal lattice\cite{Novoselov05,Wallace74,Slonczewski58,Haldane88,Castro09}. The effect of these additional
quantum numbers has already been proven to be drastically important in several quantum transport phenomena in graphene, including quantum Hall
effect\cite{Novoselov05,Zhang05,Gusynin05,Du09}, conductance quantization\cite{Peres06}, Klein tunneling\cite{Katsnelson06,Young09,Stander09}
and quantum shot noise\cite{Tworzydo06,Danneau08,DiCarlo08}. Interestingly, the pseudospin and the valley degrees of freedom in graphene have been proposed separately to be used for the controlling electronic devices in the same way as the electron spin is used in spintronic and quantum computing. Rycerz {\it et al.}\cite{Rycerz07,Xiao07,Akhmerov08,Wu11}
demonstrated an electrostatically controlled valley filter effect in graphene nanoribbons with zigzag edge which can be used for realizing valley
valve structures in valleytronics (valley-based electronics) applications. On the other hand, a pseudospin-based version of
a spin valve has been proposed in monolayer graphene and bilayer graphene\cite{Jose09,xia10,majidi11}, which can be used for realizing pseudospintronics in graphene. Also, the possibility of an interaction driven spontaneous breaking of the pseudospin symmetry, which can lead to the
realization of pseudomagnetic states in monolayer and bilayer graphene, has been studied recently\cite{Min08}.
 \par
Recent experimental progresses in proximity-inducing superconductivity in graphene by fabrication of transparent contacts between a graphene monolayer and a superconductor, has provided a unique possibility to study relativistic-like superconductivity and proximity effect\cite{Heersche07,Du08,Jeong11}. Peculiarity of Andreev reflection (AR)\cite{Andreev64},
 conversion of the electron into the hole excitations at a normal metallic-superconducting (N/S) interface, has been studied in graphene-based N/S junctions by Beenakker, who demonstrated that unlike the retro AR for highly doped graphene or a N metal, the dominant process for undoped N graphene is the specular AR\cite{beenakker06,beenakker08}. In the case of a graphene ferromagnetic-superconducting (F/S) junction, the situation is dramatically different from common F/S junctions where the subgap Andreev conductance decreases with increasing the exchange energy $h$ from its value for N/S junction and vanishes for a half metallic ferromagnet with $h=\mu$, where all carriers have the same spin\cite{de Jong95}. It has been shown that for the exchange energies higher than the chemical potential $h>\mu$, a peculiar spin-resolved Andreev-Klein process at graphene F/S interface can result in an enhancement of the subgap Andreev conductance by $h$, up to the point at which the conductance at low voltages $eV\ll\Delta_S$ is larger than its value for the corresponding N/S structure\cite{Zareyan08,Asano08,Zhang08}. Also, the corresponding Andreev-Klein bound states in graphene S/F/S structure are responsible for the long-range Josephson coupling of F graphene\cite{Moghaddam08,Linder08}. Moreover, specific nonlocal proximity effect takes place in graphene-based superconducting heterostructures mediating purely by a nonlocal process known as crossed Andreev reflection (CAR) which creates a spatially entangled electron-hole pair. While in ordinary nonrelativistic systems the small value of CAR conductance is canceled by the conductance of elastic electron cotunneling (CT) process, it can be enhanced in ballistic graphene N/S/N and F/S/F structures\cite{Cayssol08,linder09}.
\par
In this paper, we study the effect of the pseudospin degree of freedom on quantum transport in gapped monolayer graphene, that presents a
pseudospin symmetry broken ferromagnet (PF), with a finite pseudospin magnetization oriented vertically to the graphene plane. The
magnitude of the pseudomagnetization (PM) depends on the chemical potential and its direction can be switched by changing the type of
doping (electron $n$ or hole $p$). Based on this observation, we propose a nonmagnetic
pseudospin valve structure (PF/N/PF) with remarkably large pseudomagnetoresistance (PMR) in analogy to the giant
magnetoresistance (GMR) in magnetic multilayers\cite{Baibich88}, which can be perfect for chemical potentials close to
the energy gap ($\mu\simeq\Delta$) and appropriate lengths of the N region. More importantly, we show that the perfect pseudospin valve
effect can be reached even in higher chemical potentials $\mu\gg\Delta$ by applying an appropriate bias voltage. We further
demonstrate the unusually long-range penetration of the equilibrium and non-equilibrium
pseudospin polarization into the N region by proximity to a PF, that is in clear contrast to the induced magnetization
 in ordinary F/N junctions which decays exponentially within $\lambda_F$\cite{Zutic04}.
\par
Moreover, we study the effect of the pseudospin on AR and the associated proximity effect in hybrid structures of PFs and a superconductor
in S/PF and PF/S/PF geometries. We find that in graphene PF, due to the possibility for a small chemical potential $\mu$, a peculiar AR occurs at S/PF interface which is associated with an inversion of the
$z$ component of the carriers pseudospin vector, and that this has important consequences for the proximity effect. For an S/PF junction, we find that the Andreev-Klein reflection can enhance the pseudospin inverted Andreev conductance by the energy gap $\Delta_N$ to reach a limiting maximum value for $\Delta_N\gg \mu$, which depends on the bias voltage and can be larger than the value for the corresponding junction with no energy gap ($\Delta_N\ll \mu$). This is similar to the behavior of Andreev conductance with the exchange energy $h$ in a graphene F/S junction and approves that the energy gap $\Delta_N$ in the band structure of normal graphene produces an effect similar to the exchange field in F graphene. We also demonstrate that depending on the energy $\varepsilon$ of the incident electron, $\mu$ and $\Delta_N$, AR can be of retro or specular types, respectively, without or with the inversion of the $z$ component of the pseudospin vector. Furthermore, the spatially-damped oscillatory behavior of the proximity density of states in pseudoferromagnetic side of the S/PF contact and the pseudospin switching effect in superconducting graphene pseudospin valve structure (PF/S/PF) confirm the crucial rule of the pseudospin in the gapped normal graphene and its similarity to the rule of spin in an F graphene.
\par
This paper is organized as follows. We introduce pseudoferromagnets (PFs) in Sec.\ \ref{sec:level1}, and use them in Sec.\ \ref{sec:level2} to study the pseudospin valve effect in graphene PF/N/PF junction. Sections \ref{sec:level3} and \ref{sec:level4} are devoted, respectively, to the investigation of the proximity effect and the pseudospin injection in graphene PF/N junctions. In Sec.\ \ref{sec:level5}, we investigate AR in graphene S/PF junction and present our main findings for the Andreev conductance and the proximity DOS of the S/PF junction. Section \ \ref{sec:level6} is devoted to the investigation of the CT and CAR processes in superconducting pseudospin valve structure. Finally, we present the conclusion in Sec. \ \ref{sec:level7}.

\section{\label{sec:level1}Pseudoferromagnets}
One of the interesting features of graphene, that makes it very applicable in semiconductor technology, is the possibility of opening a
gap in the energy band structure of graphene. There are several methods to open an energy gap in the band structure of graphene. A
scenario is placing graphene on top of an appropriate substrate which breaks the graphene sublattice symmetry and generates a
Dirac mass for charge carriers. The band gap opening is observed in epitaxially grown graphene on a SiC
substrate\cite{Zhou07,Varchon07} and a hexagonal boron nitride crystal\cite{Giovannetti07}. The energy band gap engineering can be also achieved through doping the graphene with
several molecules such as $CrO_3$, $NH_3$, $H_2O$ \cite{Zanella08,Ribeiro08}.
\par
To study quantum transport in gapped graphene nanostructures within the scattering formalism, we first construct the
quasiparticle wave functions that participate in the scattering
processes. We adopt the Dirac equation of the form
\begin{equation}
\label{DiracFull} H_0\psi=(\varepsilon+\mu)\psi,
\end{equation}
where
\begin{equation}
\label{DiracH}
H_0=v_{F}(\bm{\sigma}\mathbf{}.\bm{p})+\Delta_N \sigma_{z}
\end{equation}
is the two-dimensional Dirac Hamiltonian in presence of an energy gap, with $\bm{p}=-i\hbar\bm{\nabla}$ the momentum operator in the $x$-$y$ plane ($v_F=10^6$ m/s represents
the Fermi velocity) and $\bm{\sigma}=(\sigma_x,\sigma_y,\sigma_z)$ the vector of the Pauli matrices operating in the space of two
sublattices of the honeycomb lattice\cite{Divincenzo84,Ando05}. The two-dimensional spinor has the form $\psi=(\psi_A,\psi_B)$,
where the two components give the amplitude of the wave function on the two sublattices and $\varepsilon$ is the quasiparticle energy.
\par
For a uniform gapped graphene region, the solutions of the Dirac equation Eq. (\ref{DiracFull}) are two states of the form
\begin{equation}
\label{psi_c}
\psi_{c}^{e\pm}=e^{\pm ik_c^{e}x} e^{iqy}
\left(
\begin{array}{c}
e^{\mp i\alpha_c^{e}/2}\\
\pm e^{-\phi_c^{e}} e^{\pm i\alpha_c^{e}/2}
\end{array}
\right),
\end{equation}
for conduction band electrons of $n$-doped graphene and
\begin{equation}
\label{psi_v}
\psi_v^{e\pm}=e^{\mp ik_v^{e}x} e^{iqy}
\left(
\begin{array}{c}
e^{\pm i\alpha_v^{e}/2}\\
\pm e^{\phi_v^{e}} e^{\mp i\alpha_v^{e}/2}
\end{array}
\right),
\end{equation}
for valance band electrons of $p$-doped graphene, at a given energy $\varepsilon$ and transverse wave vector q with the energy-momentum relation $\varepsilon_{c(v)}^{e}=\pm[-\mu+\sqrt{{\Delta_N}^2+{(\hbar v|{\bm{k}}_{c(v)}^{e}|)}^2}]$. $\alpha_{c(v)}^{e}=\arcsin[{\hbar vq}/{\sqrt{{(\varepsilon\pm\mu)}^2-{\Delta_N}^2}}]$ is the angle of propagation of electron which has longitudinal wave vector $k_{c(v)}^{e}=(\hbar v_F)^{-1}\sqrt{{(\varepsilon\pm\mu)}^2-{\Delta_N}^2}\cos\alpha_{c(v)}^{e}$ and $\phi_{c(v)}^{e}=\operatorname{arcsinh}{[\Delta_N/\sqrt{{(\varepsilon\pm\mu)}^2-{\Delta_N}^2}]}$. The two propagation directions of electron along the $x$-axis are denoted by $\pm$ in $\psi_{c(v)}^{e\pm}$.
\par
\begin{figure}
\begin{center}
\includegraphics[width=3.3in]{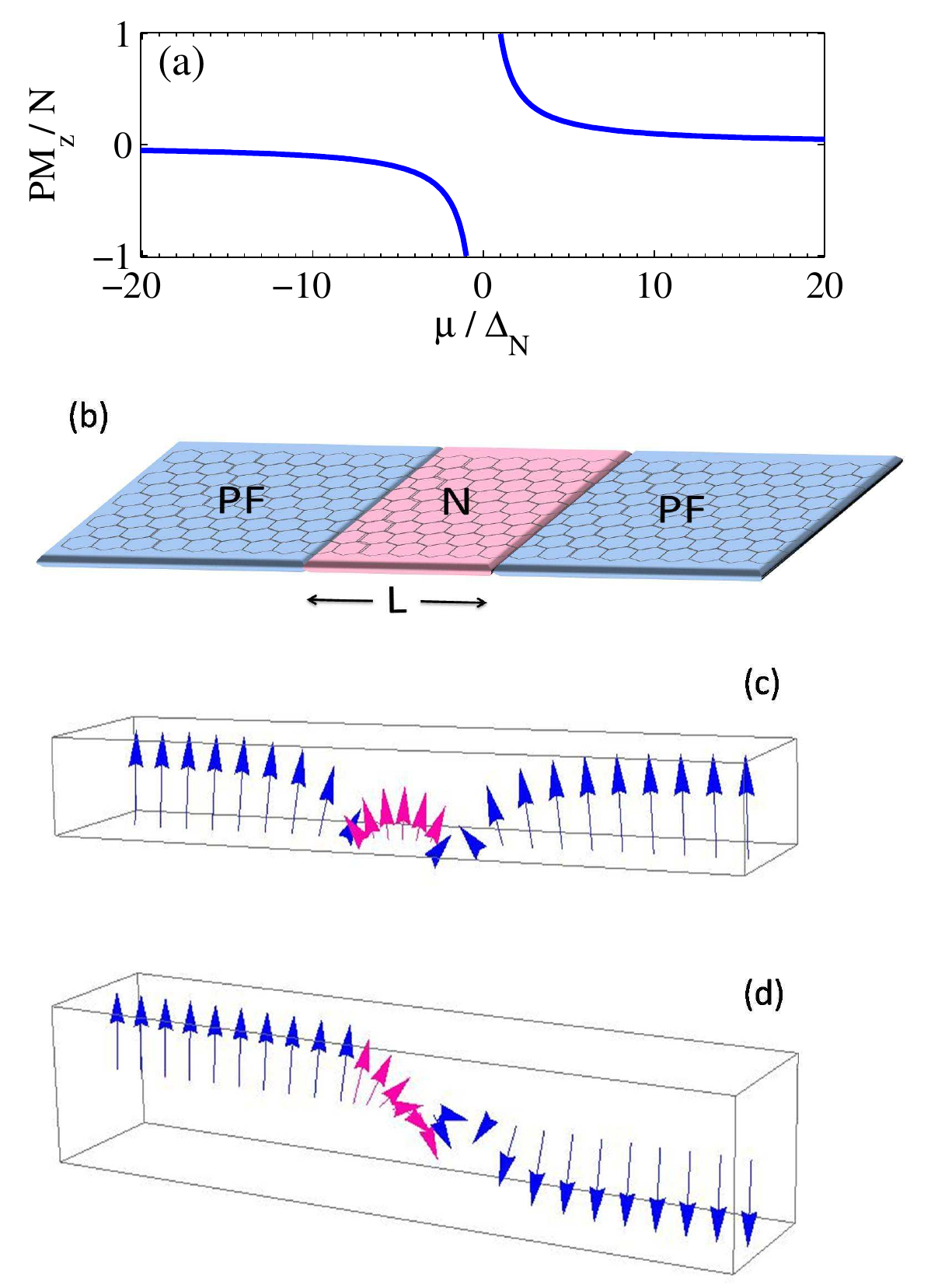}
\end{center}
\caption{\label{Fig:1}(Color online) (a) Vertical
pseudomagnetization per electron $PM_z/N$ of the gapped graphene layer
versus chemical potential $\mu$ ( $\mu$ is scaled to the energy gap $\Delta_N$). (b) Schematic illustration of the proposed
pseudospin valve in monolayer graphene: The left and right regions
are pseudoferromagnets (PFs) and the intermediate region is a normal
graphene (N) without a band gap. (c-d) Profile of pseudomagnetization vector $\bm{PM}$
inside the two PFs (blue) and the N region of
length $L=\lambda_F$ (pink) for two configurations of (c) parallel
and (d) antiparallel, when $\mu\simeq\Delta_N$.}
\end{figure}
The pseudospin of such states for conduction (valance) band electrons of $n$- ($p$-)doped graphene is obtained as
\begin{eqnarray}
\label{pseudospin}
\langle\bm{\sigma}(\bm{k})\rangle_{c(v)}^{e+}&=&\sqrt{1-{(\frac{\Delta_N}{\varepsilon\pm\mu})}^2}\ {(\cos{\alpha_{c(v)}^{e}}\ \hat{x}\pm\sin{\alpha_{c(v)}^{e}}\ \hat{y})}\nonumber\\
&+&\frac{\Delta_N}{\varepsilon\pm\mu}\ \hat{z}.
\end{eqnarray}
As can be seen from the above equation, the existence of a band gap makes the pseudospin to have a component perpendicular to the
 plane of the graphene sheet. The in-plane and out-of-plane components of the pseudospin depend on $(\varepsilon+\mu)/\Delta_N$,
which can be tuned to unity to make the pseudospin vector to be oriented perpendicular to the sheet. Increasing
$(\varepsilon+\mu)/\Delta_N$ leads to the decrease of the out-of-plane component such that it goes to zero when $\varepsilon+\mu\gg\Delta_N$.
\par
The total pseudomagnetization (PM) of the gapped graphene can be calculated by summing the expression (\ref{pseudospin}) over all the wave
vectors $\bm{k}=(k,q)$,
\begin{equation}
\label{pseudomagnetization}
\bm{PM}_{n(p)}=\sum_{\bm{k}}\langle\bm{\sigma}(\bm{k})\rangle_{c(v)}^{e+},
\end{equation}
from which we find that PM only has an out-of-plane component which depends on $(\varepsilon+\mu)/\Delta_N$. Fig. \ref{Fig:1} (a) shows the behavior of the out-of-plane component of PM per electron $PM_z/N$ as a function of $\mu/\Delta_N$ at zero temperature ($T=0$). It is seen that for $\mu\simeq\Delta_N$, $PM_z/N$ takes its maximum value $PM_z/N=1$, while increasing $\mu/\Delta_N$ leads to the decrease of $PM_z$ such that it goes to zero for highly doped gapped graphene ($\mu\gg\Delta_N$). Also the direction of PM can be switched by changing the type of doping between $n$ and $p$. We note that the pseudospin polarization of a gapped graphene corresponds to a difference in the electronic charge densities of the two triangular sublattices, which in turn produces an in-plane electrical polarization \cite{majidi11}. This correspondence between the PM vector
and the in-plane electrical polarization can be used for an experimental measuring of PM.
\par
 So we demonstrate that a monolayer graphene with an energy gap $\Delta_N$ in its electronic band structure behaves as a pseudospin symmetry-broken ferromagnet (PF)
with a perpendicular to the plane of graphene PM, whose direction is switched by altering the type of doping between $n$ and $p$.
\section{\label{sec:level2}Pseudospin valve}

Based on the above observation, we propose a nonmagnetic pseudospin valve which consists of two PFs ($x<0$ and $x>L$) with a tunable direction of PM, that are connected through a normal (nonpseudomagnetized) layer of length $L$ [shown schematically in Fig. \ref{Fig:1}(b)]. The configuration of PMs in the pseudospin valve can be changed from parallel to antiparallel by fixing the type of doping of one region and changing the type of the doping in the other region. The size of the pseudospin valve effect is determined by the extent in which the conduction of the antiparallel configuration is suppressed (similar to the spin valve effect). The pseudomagnetoresistance of a pseudospin valve is defined as
\begin{equation}
\label{PMR}
PMR=\frac{G_{P}-G_{AP}}{G_{P}},
\end{equation}
where $G_{P(AP)}$ is the conductance of the parallel (antiparallel) configuration that can be calculated from the Landauer
formula\cite{landauer88},
\begin{equation}
\label{G}
G_{P(AP)}=\frac{4 e^2}{h}\int |t_{P(AP)}|^2 \cos\alpha\ d\alpha.
\end{equation}
\begin{figure}
\begin{center}
\includegraphics[width=3in]{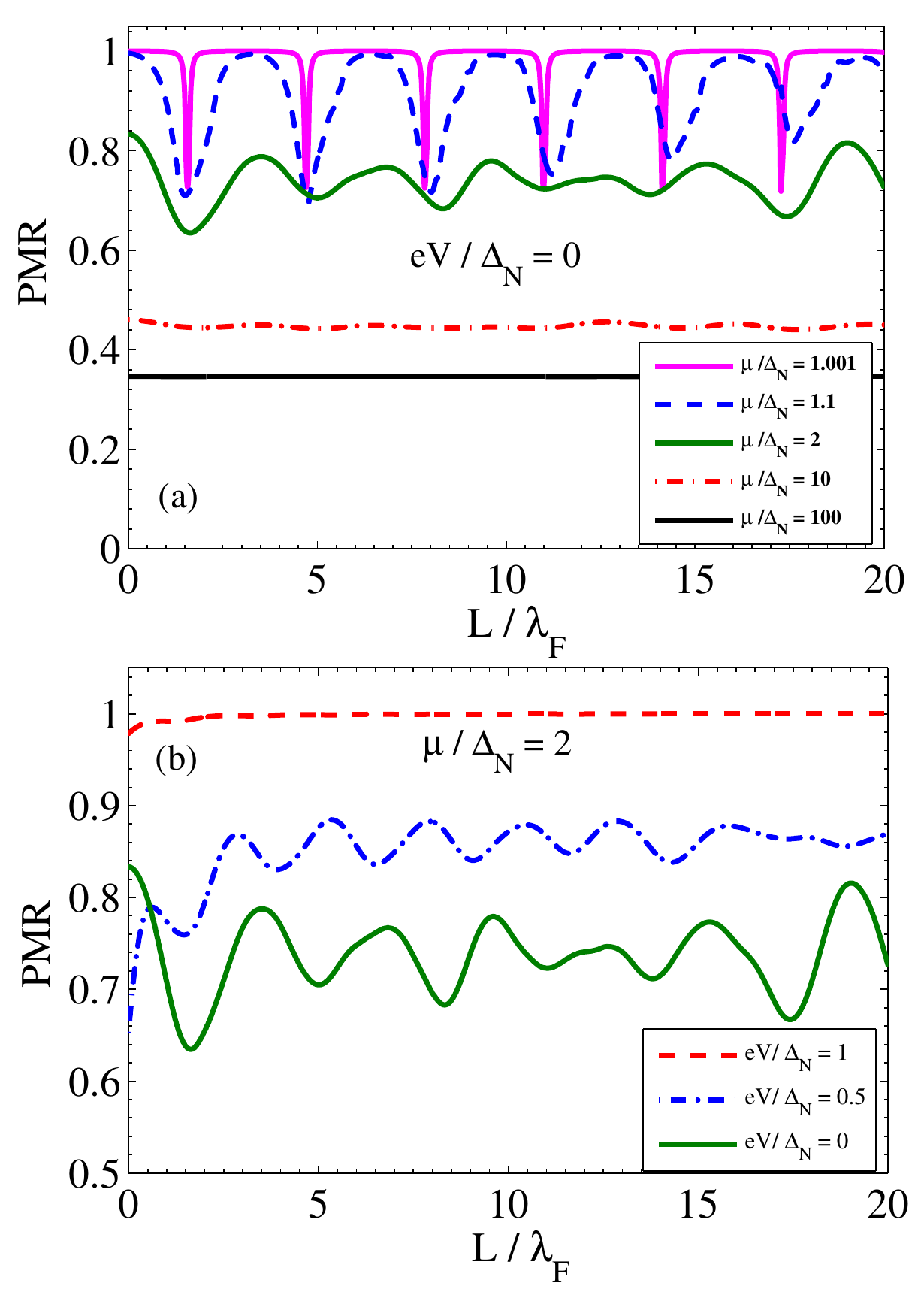}
\end{center}
\caption{\label{Fig:2}(Color online) Pseudomagnetoresistance (PMR) of
the pseudospin valve versus the length of the N region
($L/\lambda_F$) for (a) different values of $\mu/\Delta_N$, when $eV/\Delta_N= 0$, and (b) different values of the bias voltage $eV/\Delta_N$, when $\mu/\Delta_N = 2$.}
\end{figure}
Here, $t_{P(AP)}$ is the transmission amplitude of electrons through the pseudospin valve in parallel (antiparallel) configuration, which can be calculated by matching the wave functions of three regions at the two interfaces ($x=0$ and $x=L$),
\begin{eqnarray}
\label{t_p}
&&\hspace{-0.5cm}
\psi_1=\psi_c^{e+}+r\ \psi_c^{e-},\nonumber\\
&&\hspace{-0.5cm}
\psi_2=a\ {\psi'_c}^{e+}+b\ {\psi'_c}^{e-},\nonumber\\
&&\hspace{-0.5cm}
\psi_{3,P(AP)}=t_{P(AP)}\ \psi_{c(v)}^{e+}.
\end{eqnarray}

 The left PF, N region and the right PF are signed by 1,2, and 3, respectively, and $\psi_{c(v)}^{(')e\pm}$ are the wave functions of Dirac equation for incoming and outgoing electrons of $n$- ($p$-)doped graphene sheet with (without) a gap.  Figure \ref{Fig:2}(a) shows the dependence of the resulting PMR on the length of the N region $L/\lambda_{F}$ ($\lambda_F=\hbar v_F/\mu$) for different values of $\mu/\Delta_N$ at zero bias voltage $eV=0$. We have taken $\mu_1=\mu_2=|\mu_3|=\mu$. We observe that the pseudospin valve effect can be perfect ($PMR=1$) for $\mu\simeq\Delta_N$. For these values of $\mu$, PMR shows an oscillatory behavior with $L/\lambda_{F}$, with an amplitude which takes the value 1 for some ranges of the length $L$. We note that this perfect pseudospin valve effect of the monolayer graphene is more robust with respect to an increase of the length of the N region, as compared to the similar effect in a bilayer graphene pseudospin valve structure\cite{Jose09}. The amplitude of PMR decreases by increasing $\mu/\Delta_N$ and tends to the constant value of $PMR=1/3$ for a highly doped structure with $\mu\gg\Delta_N$. This residual PMR is the difference in the resistance of a $n$-$p$ graphene structure with that of a uniformly ($p$ or $n$) doped graphene with the same $|\mu|$, which is present even in the limit $PM\rightarrow 0$ of a nonpseudomagnetized structure. Also, we have found that the perfect pseudospin valve effect can be
resumed by applying an appropriate bias voltage to the valves with higher chemical potentials $\mu>\Delta_N$ [see Fig. \ref{Fig:2}(b)].

\section{\label{sec:level3}Proximity effect in PF/N junctions}
\begin{figure}
\begin{center}
\includegraphics[width=3in]{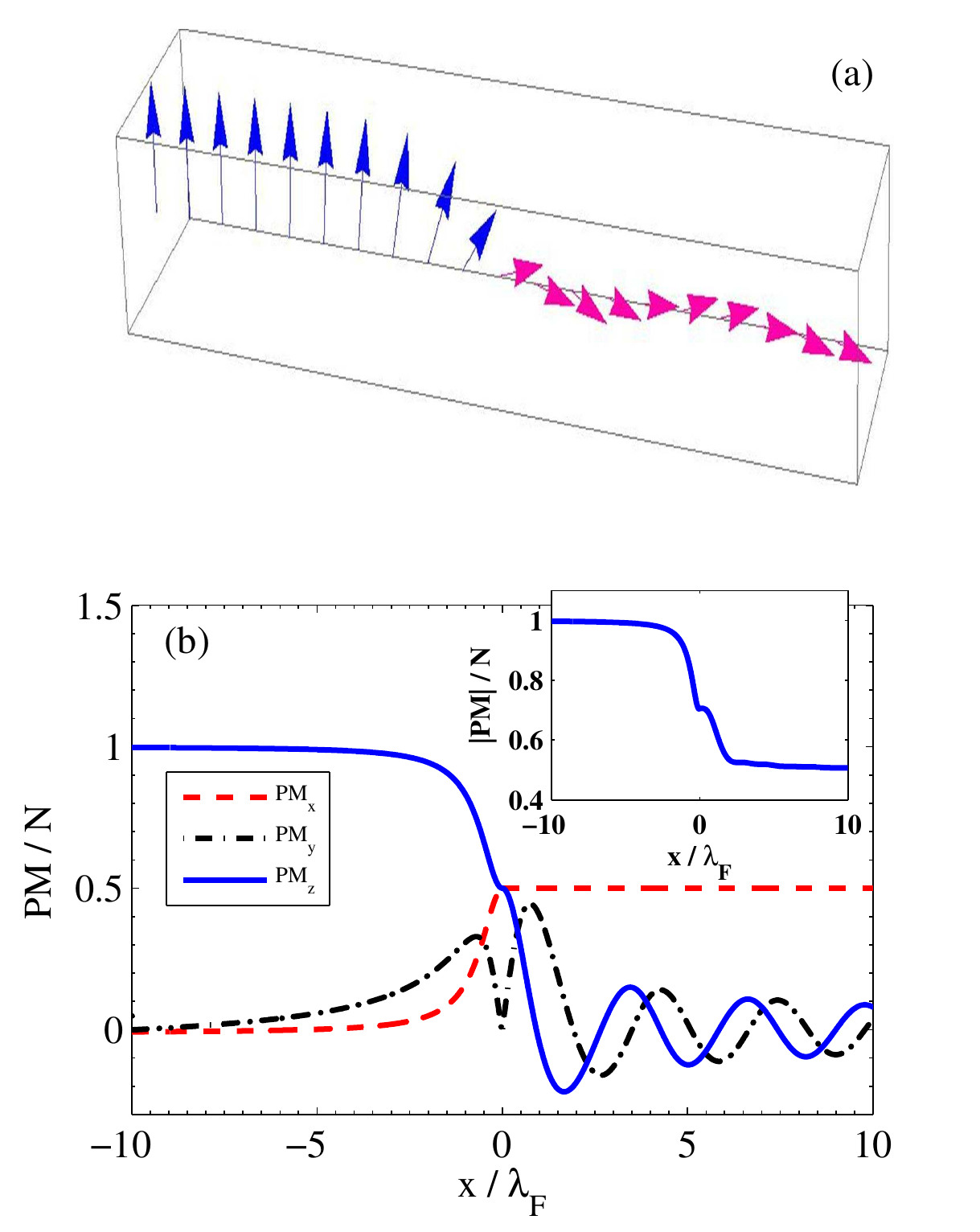}
\end{center}
\caption{\label{Fig:3}(Color online) Equilibrium PM of the PF/N junction when $\mu\simeq\Delta_N$:
(a) profile of $\bm{PM}$ in PF (blue) and N region (pink) and (b) the position dependence of the $\bm{PM}$ components.
The inset of Fig. \ref{Fig:3}(b) shows the magnitude of $\bm{PM}$.}
\end{figure}
Let us now study the proximity effect in hybrid structures of PFs and a normal graphene region in PF/N and PF/N/PF junctions.  We start with a single PF/N junction in a graphene sheet in the $x$-$y$ plane, where the region $x<0$ (PF) has a uniform PM oriented vertically to the sheet and the region $x>0$ (N region) is in the normal state. We calculate the pseudomagnetization vector $\bm{PM}$ in PF and N region using Eq. (\ref{pseudomagnetization}), and by considering the contribution of the pseudospin of all incident electrons from left ($l$) and right ($r$) regions that are scattered from the junction,
\begin{equation}
\label{PM_proximity}
\frac{\bm{PM}_i}{N_i}=\frac{1}{2}\{\frac{\bm{PM}_i^{l}}{N_i^{l}}+\frac{\bm{PM}_i^{r}}{N_i^{r}}\},
\end{equation}
where $\bm{PM}_i^{l(r)}=\sum_{\bm{k}}\langle\bm{\sigma}(\bm{k})\rangle_{\psi_{i,l(r)}}$, $N_i^{l(r)}=\sum_{\bm{k}}\psi_{i,l(r)}^*\psi_{i,l(r)}$ and $i$ denotes the PF (N region). The resulting profile of $\bm{PM}$ across the PF/N junction is demonstrated in Fig. \ref{Fig:3} for
$\mu\simeq\Delta_N$. It is seen that a nonzero $\bm{PM}$ is induced in N region ($\Delta_N=0$) which rotates around the normal to the
junction ($x$ axis) with $x$. The perpendicular component $PM_z$ oscillates as a function of $x$ with a period of order
$\lambda_F$, and shows only a weak decay in the scale of $\lambda_F$. While the in-plane components $PM_{x,y}$ vanish inside
 PF, they are produced at the PF/N interface and are penetrated into the N region. $PM_{y}$ shows an oscillatory behavior with $x$ similar to $PM_{z}$. Interestingly, $PM_x$ is uniform inside N, which considering the decay of the other two components, implies that
$\bm{PM}$ at the points in N region far from the junction is uniform and oriented perpendicular (along $x$ axis) to the $\bm{PM}$ in the connected
PF. This unusual proximity effect can be explained in terms of reflectionless Klein transmission of electrons which incident
normally to PF/N interface\cite{Katsnelson06,Young09,Stander09}. We note to the unusually long-range penetration of the proximity induced PM inside the N region, which is in contrast to the ferromagnet/normal-metal junction (F/N), in which the induced magnetization decays over
short interatomic distances.
\par
The above analysis of the proximity effect in PF/N junction can be extended to the pseudospin valve geometry of Fig. \ref{Fig:1}(b).
 The profile of $\bm{PM}$ orientation in different regions of the PF/N/PF junction is indicated in Fig. \ref{Fig:1} for parallel (c) and antiparallel (d) cases when $L=\lambda_F$ and $\mu\simeq\Delta$. $\bm{PM}$ is perpendicular to the $x$ axis and undergoes rotation across the N
contact in a way that in parallel and antiparallel cases $PM_y$ and $PM_z$, respectively, shows a change of signs at the middle of N region ($x=L/2$). Also, we obtain that the magnitude of $\bm{PM}$ inside the N region is constant with $x$ for both of parallel and antiparallel configurations. So the strong pseudomagnetic coupling between the two pseudoferromagnetic regions, which itself can be due to the long-range penetration of pseudospin polarization into the N region by proximity to PFs, leads to the strong robustness of the pseudospin valve effect with respect to increasing the length of N contact.

\section{\label{sec:level4}Pseudospin injection in PF/N junctions}
\begin{figure}
\begin{center}
\includegraphics[width=3.4in]{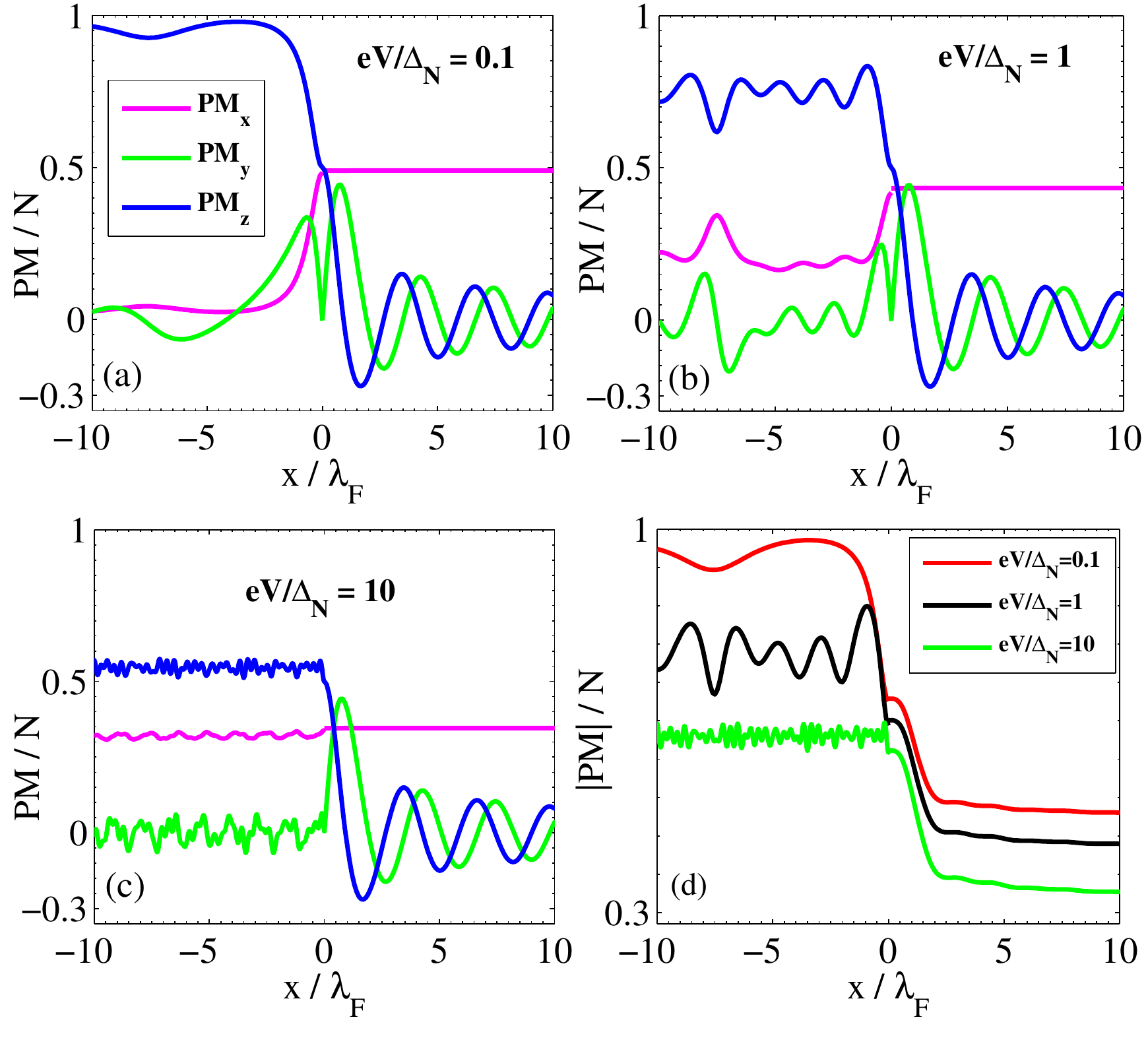}
\end{center}
\caption{\label{Fig:4}(Color online) Nonequilibrium PM of the PF/N junction when $\mu\simeq\Delta_N$: position dependence of (a-c) the $\bm{PM}$ components and (d) the magnitude of $\bm{PM}$ for three values of the bias voltage $eV/\Delta_N = 0.1, 1, 10$.}
\end{figure}
 \begin{figure}
\begin{center}
\includegraphics[width=3.4in]{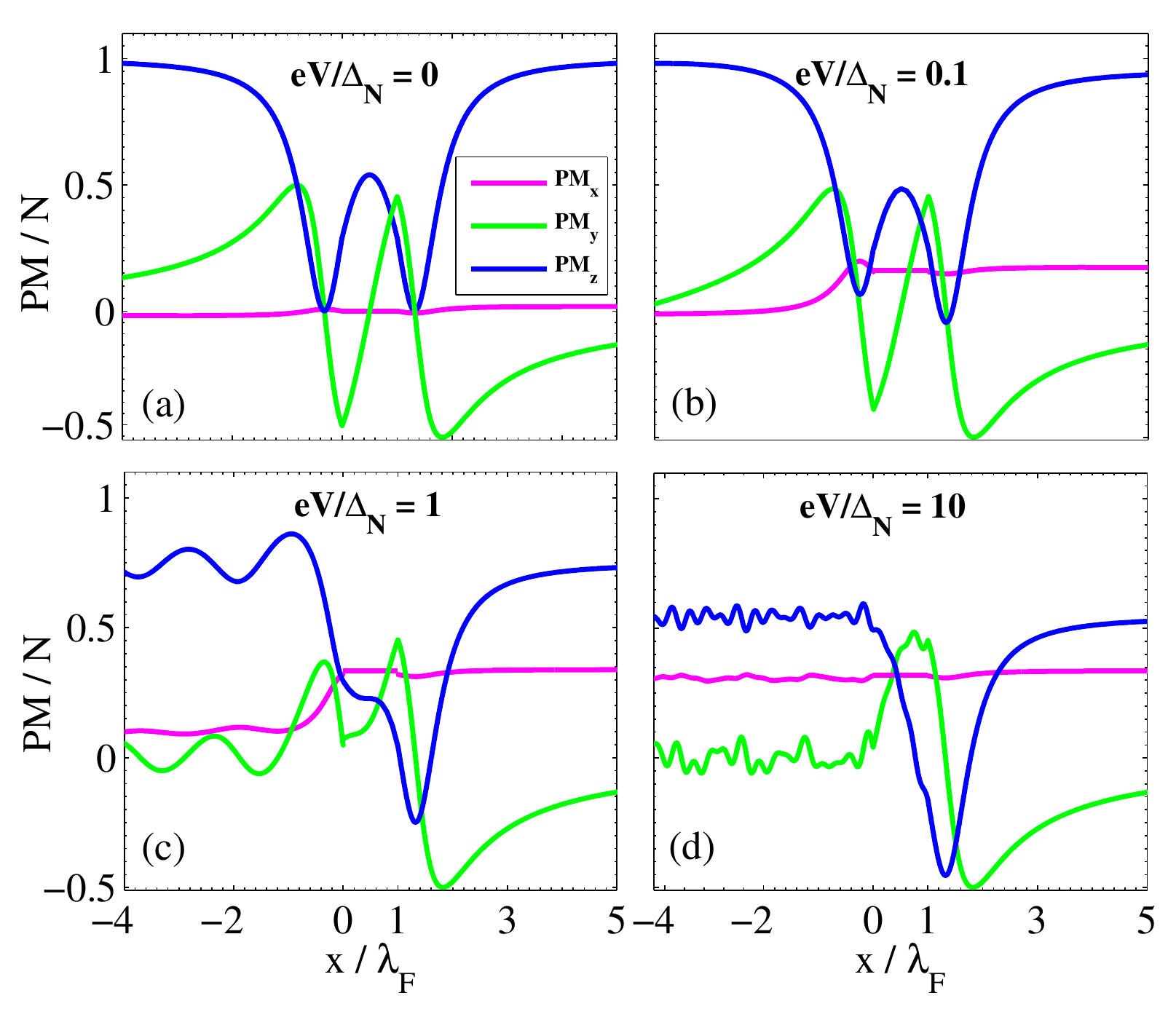}
\end{center}
\caption{\label{Fig:5}(Color online) Position dependence of the $\bm{PM}$ components of PF/N/PF junction with parallel configuration, for different values of the bias voltage $eV/\Delta_N = 0, 0.1, 1, 10$, when $L=\lambda_F$ and $\mu\simeq\Delta_N$.}
\end{figure}
In this section, we study the behavior of the injected pseudospin and pseudomagnetization into the nonpseudomagnetized N region of the PF/N junction, by the bias voltage.
In a PF/N junction, when a charge current flows across the interface, the pseudospin polarized carriers in PF contribute to the net current
 of PM entering the nonpseudomagnetized region and lead to the nonequilibrium PM in N region. Figure \ref{Fig:4} shows the behavior of the nonequilibrium PM in PF and N sides of the PF/N interface for different values of the bias voltage, when $\mu\simeq\Delta_N$. It is seen that similar to the equilibrium case, a uniform $\bm{PM}$ is induced inside the N region far from the interface, such that its magnitude decreases by increasing the bias voltage. This result is in contrast to the F/N junction, where the nonequilibrium magnetization decays exponentially within $\lambda_F$ \cite{Zutic04}. Also the magnitude of $\bm{PM}$ inside the PF decreases by increasing the bias voltage. This is due to the reduction of the $z$ component of the pseudospin vector for the charge carriers that are going away from the energy band gap, and can be seen from Figs. \ref{Fig:4}(a-c).
 \par
\begin{figure}
\begin{center}
\includegraphics[width=3in]{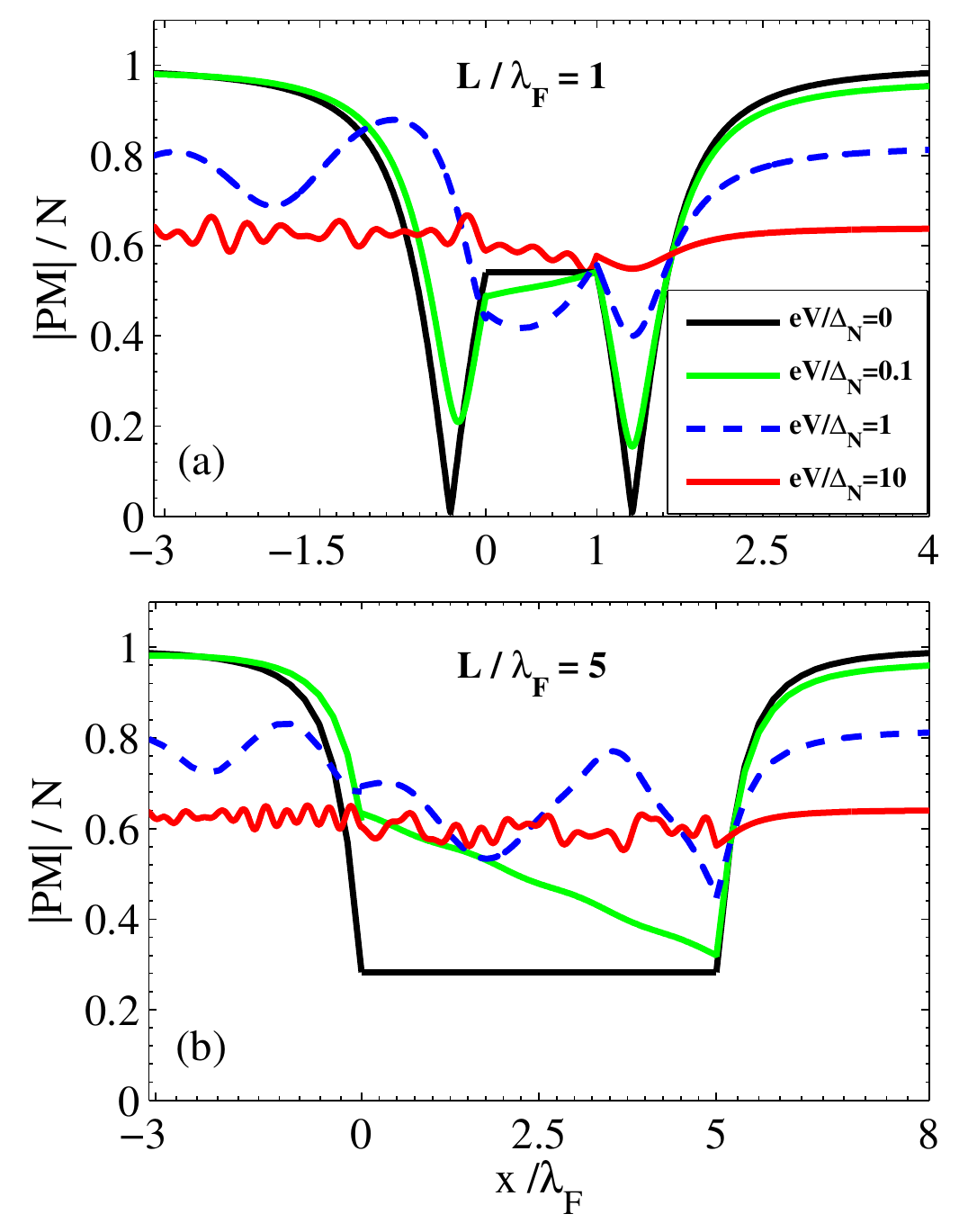}
\end{center}
\caption{\label{Fig:6}(Color online) Position dependence of the magnitude of $\bm{PM}$ for parallel configuration of the PF/N/PF junction with (a) $L=\lambda_F$ and (b) $L=5 \lambda_F$, when $eV/\Delta_N = 0, 0.1, 1, 10$ and $\mu\simeq\Delta_N$.}
\end{figure}

 The results of the above analysis for a PF/N/PF structure with parallel configuration of PMs are shown in Fig. \ref{Fig:5} for different values of the bias voltage $eV/\Delta_N = 0, 0.1, 1, 10$, when $\mu=\Delta_N$ and $L=\lambda_F$. It is seen that in contrast to the case of equilibrium, the $x$ component of the injected $\bm{PM}$ into the N region has a nonzero constant value and the symmetry of the $y$ and $z$ components $PM_{y,z}$ are broken relative to the middle of the N region. Therefore, in contrast to the constant magnitude of the induced $\bm{PM}$ in equilibrium, the magnitude of the nonequilibrium $\bm{PM}$ depends on $x$ and has a different behavior for different lengths of the N region [Fig. \ref{Fig:6}].
\section{\label{sec:level5}Andreev reflection in graphene S/PF junction}
\begin{figure}
\begin{center}
\includegraphics[width=3.3in]{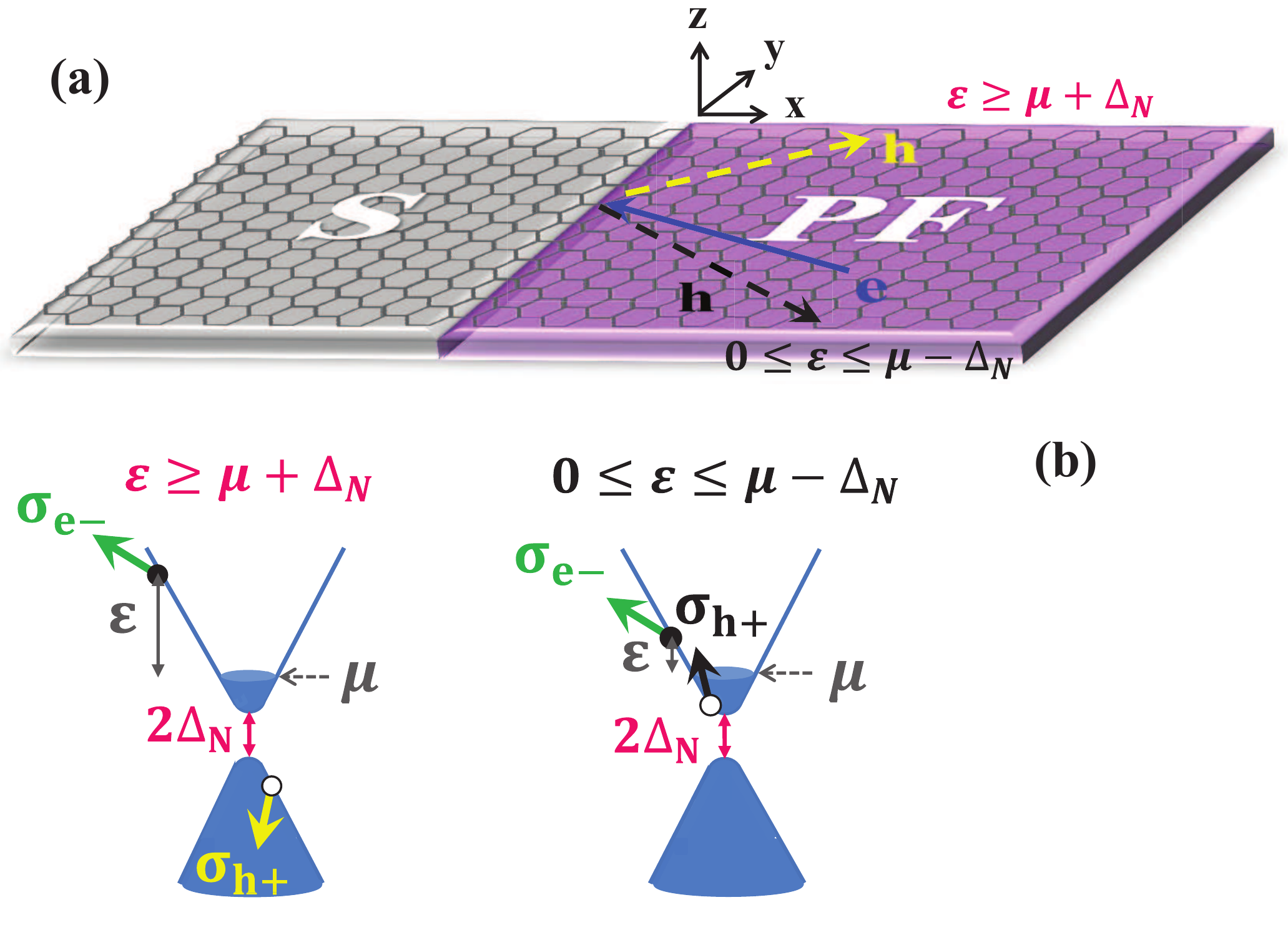}
\end{center}
\caption{\label{Fig:7}(Color online) (a) Schematic illustration of the graphene S/PF junction. (b) The band structure of the $n$-doped PF to explain the two cases of Andreev reflection at S/PF interface: Right (Left) panel shows that an incident electron from the conduction band of PF with a subgap energy $0\leq\varepsilon\leq \mu-\Delta_N$ ($\varepsilon\geq \mu+\Delta_N$) is retro (specular) reflected as a hole in the conduction  (valance)band without (with) the inversion of the $z$ component of the pseudospin vector at S/PF interface. $\bm{\sigma}_{e-}$ and $\bm{\sigma}_{h+}$ denote the pseudospin vectors of incident electron and reflected hole. $v_{e}$ and $v_{h}$ denote the velocity vectors of the electron and the hole, moving in different directions.}
\end{figure}

Now, we consider a wide graphene S/PF junction normal to $x$-axis with highly doped S region for $x<0$ and $n$-doped PF for $x>0$ [see Fig. \ref{Fig:7}(a)]. The S region can be produced by depositing S electrode on top of the graphene sheet\cite{Heersche07,Du08,Jeong11}.
In this region $\Delta_N=0$ and the superconducting correlations are characterized by the superconducting pair potential (order parameter) $\Delta_S$ which is taken to be real and constant. To study AR at S/PF interface within the scattering formalism, we first construct the quasiparticle wave functions that participate in the scattering processes. In order to describe the superconducting correlations between relativistic electrons and holes of different valleys, we adopt the Dirac-Bogoliubov-de Gennes (DBdG) equation:\cite{beenakker06}
\begin{equation}
\label{DBdG}
\hspace{-0.5cm}\left(
\begin{array}{cc}
H-\mu & \Delta_S \\
\Delta_{S}^{\ast}& \mu-H
\\
\end{array}
\right)
\left(
\begin{array}{c}
u\\
v
\end{array}
\right)
=\varepsilon\left(
\begin{array}{c}
u\\
v
\end{array}
\right),
\end{equation}
\begin{equation}
\label{H}
H=H_0-U(\bm{r}),
\end{equation}
where $H_0$ is the two-dimensional Dirac Hamiltonian with an energy gap (Eq. (\ref{DiracH})), $\varepsilon$ is the excitation energy and $U(\bm{r})$ the electrostatic potential is taken to be $U_0\gg \mu$ in S region and $U=0$ in PF. The electron and the hole wave functions, $u$ and $v$, are two-component spinors of the form $(\psi_A,\psi_B)$.
\par
 An incident electron of the conduction band from right to S/PF interface with a subgap energy $\varepsilon\leq\Delta_S$ can be either normally reflected as an electron or Andreev reflected as a hole. The reflected hole can be from the conduction or the valance band, depending on the electron energy $\varepsilon$, $\mu$ and the energy gap $\Delta_N$. As is shown in Fig. \ref{Fig:7}(b), as long as $0\leq\varepsilon\leq \mu-\Delta_N$ the reflected hole is an empty state in the conduction band and AR is retro (middle panel), while for $\varepsilon\geq \mu+\Delta_N$ it is an empty state in the valance band and AR is specular, if $\Delta_N<\Delta_S$ (right panel). The importance of AR near the Fermi level imposes the condition of $\Delta_N<\Delta_S$ on size of the energy gap $\Delta_N$. The retro reflection dominates if $\mu\gg\Delta_S+\Delta_N$, while the specular reflection dominates if $\mu\ll\Delta_S-\Delta_N$. Using the solutions of Dirac equation for electrons and holes of $n$-doped PF, the pseudospin of the incident electron and the reflected hole of the conduction (valance)band are obtained as,
\begin{eqnarray}
\label{pseudospin_e}
&&\hspace{-5mm}\langle\bm{\sigma}(\bm{k})\rangle_{c}^{e-}=\sqrt{1-{(\frac{\Delta_N}{\mu+\varepsilon})}^2}\ (-\cos{\alpha_{c}^e}\ \hat{x}+\sin{\alpha_{c}^e}\ \hat{y})\nonumber\\
&&\hspace{1.4cm}+\frac{\Delta_N}{\mu+\varepsilon}\ \hat{z},\\
\label{pseudospin_h}
&&\hspace{-5mm}\langle\bm{\sigma}(\bm{k})\rangle_{c(v)}^{h+}=\sqrt{1-{(\frac{\Delta_N}{\mu-\varepsilon})}^2}\ (-\cos{\alpha_h}\ \hat{x}\pm\sin{\alpha_h}\ \hat{y})\nonumber\\
&&\hspace{1.5cm}\pm\frac{\Delta_N}{|\mu-\varepsilon|}\ \hat{z}.
\end{eqnarray}
Here, $\alpha_{h}=\arcsin{[{\hbar vq}/{\sqrt{(\mu-\varepsilon)^2-{\Delta_N}^2}}]}$ indicates the angle of propagation of the hole at a transverse momentum $q$ with energy-momentum relation $\varepsilon_{c(v)}^h=\mu\mp\sqrt{{\Delta_N}^2+(\hbar v|{\bm{k}}_{h}|)^2}$  for the hole from the conduction (valance)band. As can be seen from the above equations when an electron from the conduction band is reflected as a hole in the valance band, the sign of the gap-induced $z$ component of the pseudospin vector $\langle{\sigma}_z\rangle$ is changed, while in the case of the conduction band hole, it retains its sign.
 This is shown schematically in the middle (right) panel of Fig. \ref{Fig:7}(b) for the case of Andreev reflected hole from the conduction  (valance)band without (with) the inversion of $\langle{\sigma}_z\rangle$ upon AR at S/PF interface.
 Thus, for the incident electron and the reflected hole being from different types of bands, we have an inversion of the $z$ component of the pseudospin vector upon AR at S/PF interface. In the following we will show how the pseudospin $\langle{\sigma}_z\rangle$ inversion by AR leads to peculiar properties of S/PF and PF/S/PF systems.
\par
To evaluate the Andreev conductance of an S/PF junction, we use the Blonder-Tinkham-Klapwijk (BTK) formula\cite{Blonder82}:
\begin{equation}
\label{G}
G_{c(v)}=\frac{4e^2}{h}\tilde{N}(eV)\int_{0}^{\alpha_{c}}(1-|r_{c(v)}|^2+|r_{A,c(v)}|^2)\cos\alpha_e\ d\alpha_e,\\
\end{equation}
where $r_{c(v)}$ and $r_{A,c(v)}$ denote the amplitudes of normal and Andreev reflections, respectively. $\tilde{N}(\varepsilon)={W{(\mu+\varepsilon)}^2}/{\pi\hbar v_F\sqrt{{(\mu+\varepsilon)}^2-{\Delta_N}^2}}$ is the number of transverse modes in a sheet of gapped graphene of width W and $\alpha_{c}=\arcsin[\sqrt{{{(\mu-\varepsilon)}^2-{\Delta_N}^2}}/\sqrt{{{(\mu+\varepsilon)}^2-{\Delta_N}^2}}]$
is the critical angle of incidence above which the Andreev reflected waves become evanescent and do not contribute to any transport of charge.
\par
We calculate the amplitudes of normal and Andreev reflections by matching the wave functions of PF and S region at the interface $x=0$. The wave functions inside PF and S region are as follows:
\begin{eqnarray}
\label{pf wave function}
&&\psi_{PF}^{c(v)}=\psi_{c}^{e-}+r_{c(v)}\ \psi_c^{e+}+r_{A,c(v)}\ \psi_{c(v)}^{h+},\\
\label{s wave function}
&&\psi_{S}=a\ \psi^{S+}+b\ \psi^{S-},
\end{eqnarray}
where $\psi_{c(v)}^{e(h)\pm}$ and $\psi^{S\pm}$ are the solutions of DBdG equation for the quasiparticles inside the $n$-doped PF and S region, respectively, and the two cases of the Andreev reflected holes from the conduction (valance)band without (with) the inversion of $\langle{\sigma}_z\rangle$ are denoted by $c(v)$ in $\psi_{PF}^{c(v)}$.

\begin{figure}
\begin{center}
\includegraphics[width=3.3in]{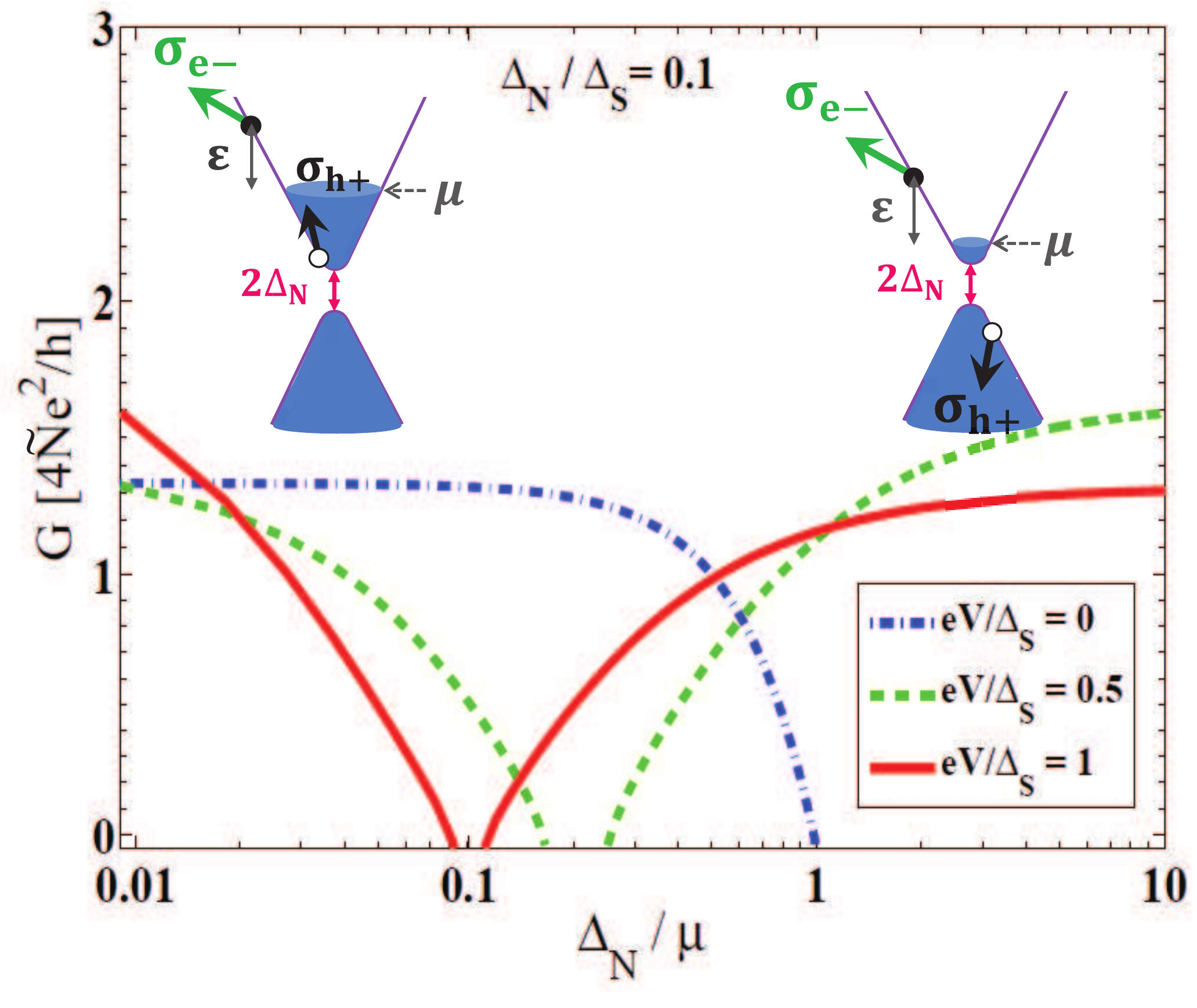}
\end{center}
\caption{\label{Fig:8}(Color online) Dependence of the Andreev conductance of graphene S/PF contact on the gap $\Delta_N/\mu$ (in units of the chemical potential) at three bias voltages $eV/\Delta_S = 0,0.5,1$. Left (Right) inset shows that an incident electron from the conduction band of PF with ${\Delta_N/\mu}\leq{1/(1+eV/\Delta_N)}$ (${\Delta_N/\mu}\geq{1/(eV/\Delta_N-1)}$) is reflected as a hole in the conduction (valance)band without (with) the inversion of the $z$ component of the pseudospin vector at S/PF interface.}
\end{figure}
\par
Figure \ref{Fig:8} shows the Andreev conductance of the graphene S/PF junction as a function of $\Delta_N/\mu$ for $\Delta_N/\Delta_S=0.1$ and different subgap bias voltages. It is seen that for $\Delta_N<{\mu/(1+eV/\Delta_N)}$, the conductance decreases monotonically with $\Delta_N/\mu$. In this interval, the incident electron and the reflected hole are from the conduction band and therefore AR is without the inversion of $\langle{\sigma}_z\rangle$ [see left inset of Fig. \ref{Fig:8}]. The density of states of the conduction band hole decreases by increasing $\Delta_N/\mu$. Thus, the amplitude of AR and hence the Andreev conductance decreases with $\Delta_N/\mu$ and goes to zero at $\Delta_N={\mu/(1+eV/\Delta_N)}$, where the density of states of the conduction band hole vanishes. The absence of hole states for ${\mu/(1+eV/\Delta_N)}<\Delta_N<{\mu/(eV/\Delta_N-1)}$ causes a gap in conductance, which decreases with $eV/\Delta_S$ and goes towards smaller $\Delta_N/\mu$. For $\Delta_N\geq{\mu/(eV/\Delta_N-1)}$, the pseudospin $\langle{\sigma}_z\rangle$ inverted Andreev conductance increases monotonically with $\Delta_N/\mu$. In this regime, the transport is between the conduction and the valance band and the incident electron of the conduction band is reflected as a hole in the valance band. So the pseudospin $\langle{\sigma}_z\rangle$ of the reflected hole changes sign [see right inset of Fig. \ref{Fig:8}] and the density of states of the hole increases with $\Delta_N/\mu$, resulting in an enhancing Andreev conductance. Such a  peculiar AR is associated with a Klein tunneling of the $n$-type carriers to the $p$-type carriers. The enhancing conductance reaches a limiting maximum value for $\Delta_N\gg \mu$, which depends on the bias voltage and can be larger than the value for the corresponding S/N structure ($\Delta_N\ll \mu$). The limiting value of the Andreev conductance for $\Delta_N\gg \mu$ decreases by increasing $\Delta_N/\Delta_S$ from its value for a specular AR in corresponding S/N structure $\Delta_N\ll\Delta_S$ $(G/G_0 = 4/3)$ and vanishes for $\Delta_N>\Delta_S$, while for $\Delta_N\ll \mu$ it increases by increasing $\Delta_N/\Delta_S$ and tends to the corresponding value of a retro type AR $(G/G_0 = 2)$ as $\Delta_N\rightarrow\Delta_S$ \cite{majidi12}.
 So the behavior of the Andreev conductance with $\Delta_N/\mu$ is similar to that of a graphene F/S junction with $h/\mu$, where AR of $n$-$n$ type carriers for $h<\mu$ changes to the Andreev-Klein reflection of the $n$-$p$ type carriers for $h>\mu$\cite{Zareyan08}. This shows that the energy gap $\Delta_N$ in the band structure of normal graphene behaves like an exchange energy in F graphene and enhances
 the subgap Andreev conductance of S/PF junction, which is accompanied by the inversion of the $z$ component of the pseudospin vector for the reflected hole relative to the incident electron.
 \par
To complete the analysis of the present section, we evaluate the proximity density of states (DOS) inside the pseudoferromagnetic region by using the formula \cite{Gennes89}
\begin{equation}
\label{N}
N(\varepsilon,r)=\sum_{\bm{k}}{|\psi_{\bm{k}}(r)|^2\ \delta(\varepsilon(\bm{k})-\varepsilon)},\\
\end{equation}
where $\psi_{\bm{k}}(r)$ corresponds to the eigenfunction of energy $\varepsilon(\bm{k})$ and the sum is over all states with the wave vectors  $\bm{k}$. Replacing Eq. (\ref{pf wave function}) in the above equation, we find the total subgap DOS inside the pseudoferromagnetic region for the case of AR with $\langle{\sigma}_z\rangle$ inversion as,
\begin{eqnarray}
&&\hspace{-20mm}\frac{N(\varepsilon, x)}{N_{0}(\varepsilon)} =\frac{1}{4}\sqrt{1- (\frac{\Delta_N}{\mu+\varepsilon})^2}\nonumber\\
&&\int_{-\pi/2}^{\pi/2} |{\psi(r)}_{PF}^{v}|^2\ {\cos}^2\alpha_e\ d\alpha_e,
\end{eqnarray}
where $N_0(\varepsilon)= {(\mu+\varepsilon)^2}/{(\pi\hbar v_F)^2 \sqrt{(\mu+\varepsilon)^2-{\Delta_N}^2}}$ is the DOS of a pseudoferromagnetic layer.
\begin{figure}
\begin{center}
\includegraphics[width=3.3in]{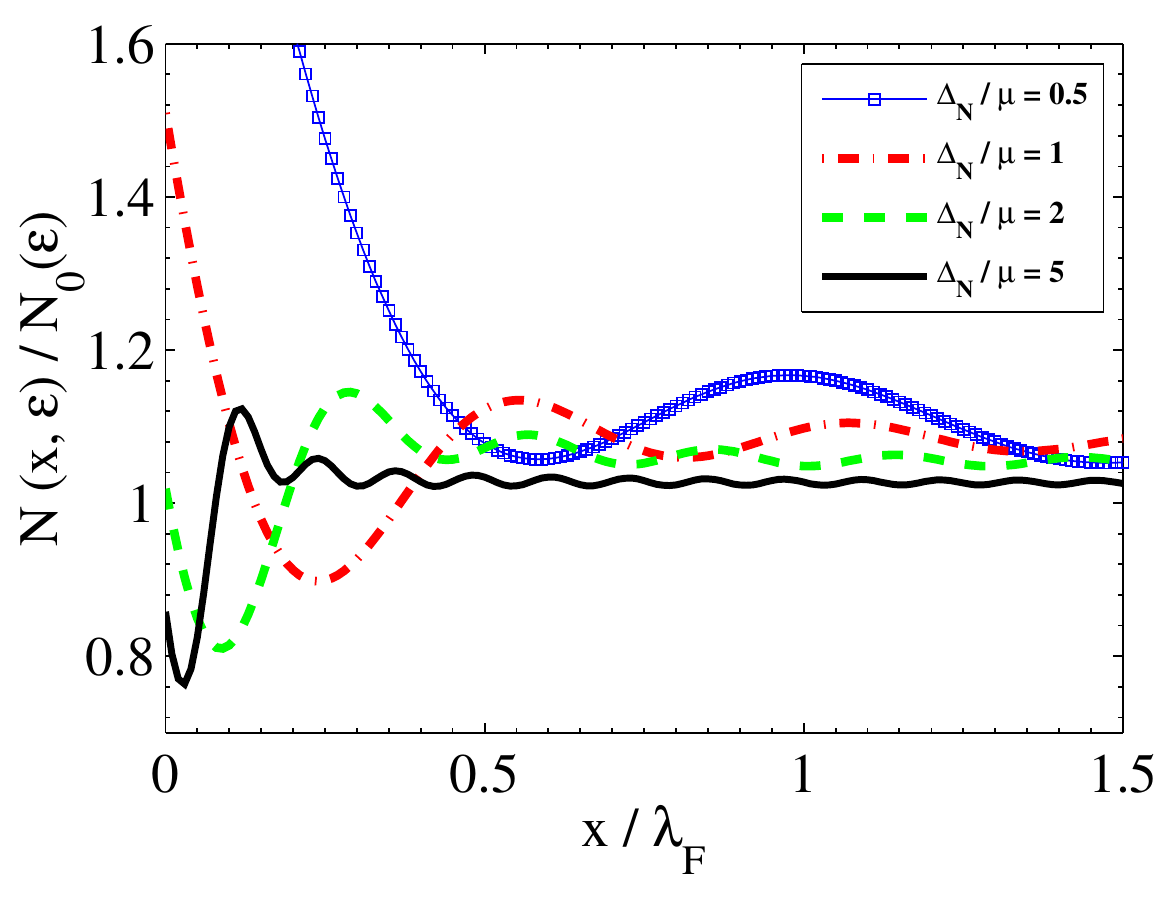}
\end{center}
\caption{\label{Fig:9}(Color online) The behavior of the proximity density of states (DOS) inside the pseudoferromagnetic region versus $x/\lambda_F$ for different values of $\Delta_N/\mu$, when $\varepsilon/\Delta_S = 0.5$ and $\Delta_N/\Delta_S = 0.1$. }
\end{figure}
Figure \ref{Fig:9} shows the behavior of the proximity DOS inside the pseudoferromagnetic region in terms of the dimensionless distance $x/\lambda_F$ for different values of $\Delta_N/\mu$, when $\varepsilon/\Delta_S = 0.5$ and $\Delta_N/\Delta_S = 0.1 $. We see that there are two phenomena to consider in describing the spatial variations of $N(x)$, when an energy gap is present in the band structure of normal graphene. The first phenomenon is the short distance decay at the interface with a slope which increases by increasing $\Delta_N/\mu$. The other important phenomenon is the damped oscillation of $N(x)$, caused by the momentum shift between Andreev correlated electron-hole pair with opposite $\langle{\sigma}_z\rangle$ directions. The period of oscillations is determined by $\hbar v_F/\Delta_N$, which is similar to an S/F structure where the period of DOS oscillations in the ballistic limit is given $\hbar v_F/h$ \cite{Zareyan01,Zareyan02,Buzdin05}. This shows the similarity of the effect of an spin-splitting exchange field $h$ with the energy gap $\Delta_N$, which behaves as a pseudospin-splitting field [see Eq. (\ref{DiracH})]. We note that the appropriate method to probe the DOS oscillations in S/PF junction is the local scanning of the surface of the pseudoferromagnetic region by scanning tunneling microscopy (STM). So the spatially damped oscillatory behavior of the DOS inside the pseudoferromagnetic region confirms that the energy gap $\Delta_N$ in the band structure of normal graphene
produces an effect similar to the exchange field in F graphene.
\section{\label{sec:level6}Superconducting pseudospin valve}
Finally, we study the nonlocal quantum transport in PF/S/PF junction that constitutes a superconducting pseudospin valve structure. We calculate the normal and Andreev reflection amplitudes ($r$ and $r_A$, respectively) in the left PF and the transmission amplitudes of the electron ($t$) and the hole ($t_A$) into the right PF of both parallel and antiparallel configurations, by matching the wave functions of the two PFs and S region at the two interfaces ($x=0$ and $x=L$),
\begin{eqnarray}
\label{pfspf wave function}
&&\psi'_{1}=\psi_{c}^{e+}+r\ \psi_c^{e-}+r_{A}\ \psi_{c}^{h-},\nonumber\\
&&\psi'_{2}=a\ \psi^{S+}+b\ \psi^{S-}+a'\ \psi^{S'+}+b'\ \psi^{S'-}\nonumber,\\
&&\psi'_{3,P(AP)}=t\ \psi_{c(v)}^{e+}+t_{A}\ \psi_{c(v)}^{(')h+}.
\end{eqnarray}
Here, the left PF, S region, and the right PF are signed by 1,2, and 3, respectively, and $\psi_{c(v)}^{(')h+}$ is the solution of the Dirac equation for conduction (valance)band hole of the $n$- ($p$-)doped PF). Replacing the reflection and transmission amplitudes in BTK  formula, we obtain the conductance of AR, CT, and CAR processes for parallel and antiparallel alignments of PMs. In CAR process an electron excitation and a hole excitation at two separate pseudoferromagnetic leads are coupled by means of Andreev scattering processes at two spatially distinct interfaces. We find that for all incoming waves with two bias voltages $eV = \pm (\mu-\Delta_N)$, AR process is suppressed and the cross-conductance in the right PF depends crucially on the configuration of PMs in the two PFs. We find that the transport is mediated purely by CT in parallel configuration and changes to the pure CAR in the low energy regime, by reversing the direction of PM in the right PF. This suggests a pseudospin switching effect between the pure CT and pure CAR in PF/S/PF structure, which can be seen from Eq. (\ref{pseudospin_PF/S/PF}), for the right going conduction (valance)band electron (hole) of $n$- ($p$-)doped PF,
\begin{eqnarray}
\label{pseudospin_PF/S/PF}
&&\hspace{-5mm}\langle\bm{\sigma}(\bm{k})\rangle_{c(v)}^{e(h)+}=\pm\sqrt{1-{(\frac{\Delta_N}{\mu+\varepsilon})}^2}\ (\cos{\alpha_{c}^e}\ \hat{x}+\sin{\alpha_{c}^e}\ \hat{y})\nonumber\\
&&\hspace{1.4cm}\pm\frac{\Delta_N}{\mu+\varepsilon}\ \hat{z}.
\end{eqnarray}
\par
\begin{figure}
\begin{center}
\includegraphics[width=3.4in]{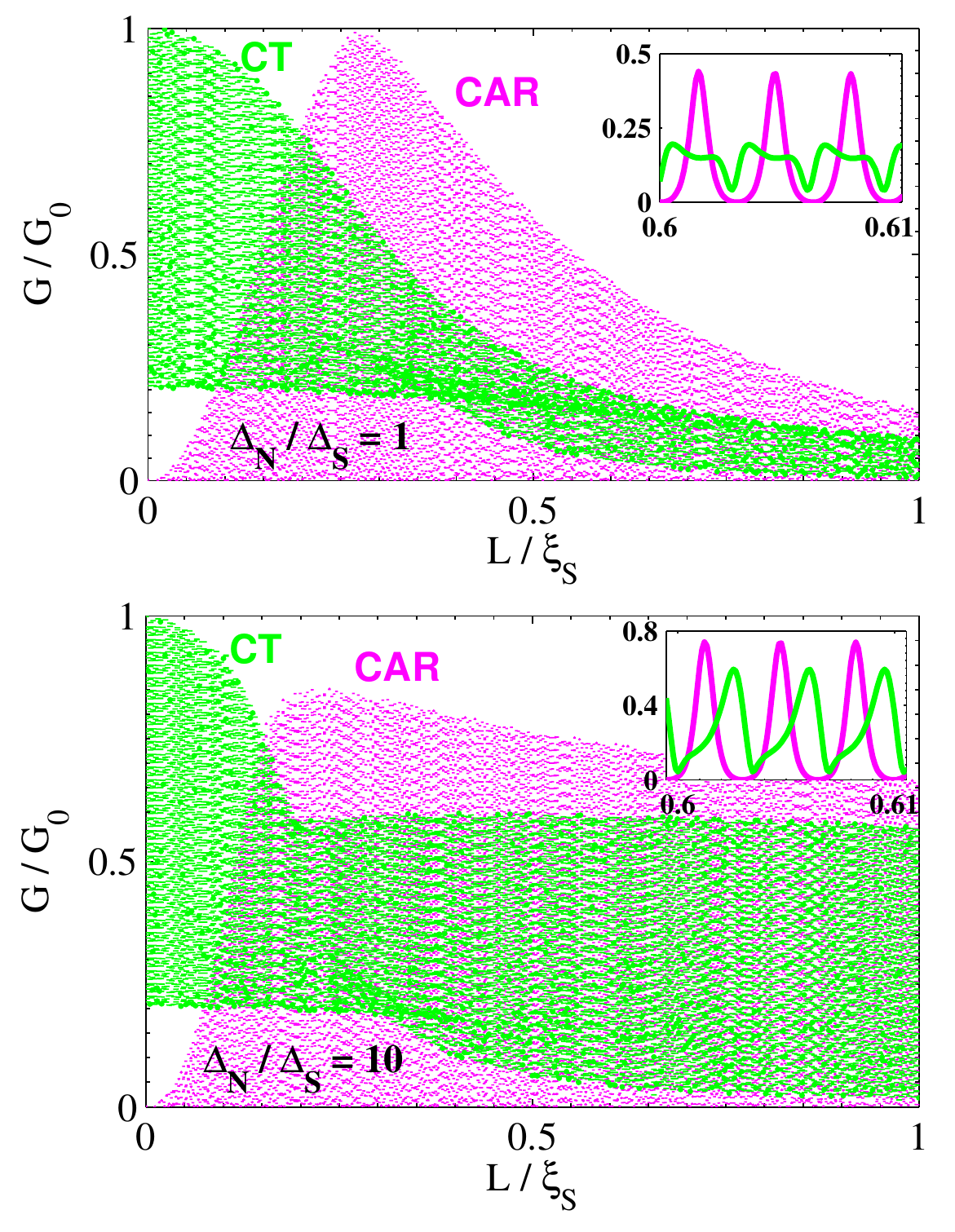}
\end{center}
\caption{\label{Fig:10}(Color online) Plots of the conductance for CT and CAR processes, respectively, in parallel ($G_{CAR}\rightarrow 0$) and antiparallel ($G_{CT}\rightarrow 0$) alignments of PMs versus the length of the S region for two values of $\Delta_N/\Delta_S=1,10$, when $\mu/\Delta_N= 1.1$ and $eV = \mu - \Delta_N$.}
\end{figure}
Figure \ref{Fig:10} shows the behavior of the conductance of CT and CAR processes, respectively, in parallel and antiparallel alignments of PMs versus the length of the S region for two values of $\Delta_N/\Delta_S$, when $\mu / \Delta_N = 1.1$ and $eV = \mu - \Delta_N$. It is seen that the CT process is favored for short junctions $L\ll \xi_{s}$, while the CAR process is suppressed in this regime. The CT conductance drops by increasing the length $L$, while the CAR conductance peaks at $L<\xi_{s}$. We see that the conductance of CT and CAR processes have oscillatory behavior with $L / \xi_{s}$ and increase by increasing $\Delta_N/\Delta_S$ from their values for the corresponding graphene N/S/N structure. Also we can see that in contrast to the graphene N/S/N structure, CT and CAR processes are present for long lengths of the S region, respectively, in parallel and antiparallel PM configurations.
This effect is similar to graphene F/S/F structure \cite{linder09} and shows that the gapped normal graphene behaves like an F graphene.
\section{\label{sec:level7}Conclusion}

In conclusion, we have demonstrated the unusual features of the pseudospin polarized quantum transport in graphene-based hybrid structures of normal (N) regions, superconductors (S) and gapped regions as pseudoferromagnets (PFs). A gapped graphene is in a sublattice pseudospin symmetry-broken state with a net pseudomagnetization (PM) oriented perpendicularly to the plane of graphene. The magnitude of PM depends on the ratio of the chemical potential to the energy gap $\mu/\Delta_N$ and its direction is switched by changing the type of doping between $n$ and $p$. Based on this observation, we have proposed a perfect pseudospin valve (PF/N/PF junction) with pseudomagnetoresistance $PMR=1$, for $\mu\simeq\Delta_N$ and appropriate contact length $L$, whose magnetization alignments can be controlled by altering the type of their doping. We have shown that this perfect pseudomagnetic valve effect is preserved even for very large lengths $L\gg \lambda_F$. Also, it can be resumed at large chemical potentials by
applying an appropriate bias voltage. We have explained this strong robustness of the perfect pseudomagnetic switching with respect to increasing of the contact length, in terms of an unusually long-range penetration of an equilibrium and nonequilibrium pseudospin polarization into the normal region by proximity to a PF. The induced pseudomagnetization vector $\bm{PM}$ undergoes a damped spatial precession  around the normal to the PF/N junction and tends to be uniform along the normal at the large distances $x\gg \lambda_F$ from the junction.
\par
Furthermore, we have found that upon a certain condition, Andreev reflection (AR) of an electron from an S/PF interface is associated with an inversion of the perpendicular component of its pseudospin, and that this has important consequences for the proximity effect in S/PF and PF/S/PF geometries. For an S/PF junction system, we have found that the Andreev-Klein reflection can enhance the amplitude of AR and the resulting Andreev conductance by $\Delta_N$. In particular, we have shown that depending on the bias voltage the Andreev conductance of weekly doped PF ($\mu\ll\Delta_N$) can be larger than its value for the corresponding graphene S/N junction. This is similar to the behavior of Andreev conductance with the exchange energy $h$ in a graphene ferromagnet-superconductor junction. We have further studied the proximity density of states (DOS) in pseudoferromagnetic side of the S/PF contact, which exhibit a damped-oscillatory behavior as a function of the distance from the interface. The period of DOS oscillations is found to be inversely proportional to the energy gap $\Delta_N$. The proximity DOS in ferromagnetic graphene shows similar spatial oscillations with a period determined by $1/h$. For a superconducting pseudospin valve (PF/S/PF) structure, we have found that the transport is mediated purely by elastic electron cotunneling process in parallel alignment of PMs and crossed Andreev reflection process in antiparallel configuration, that is accompanied by pseudospin switching effect. This is again similar to the behavior of the corresponding superconducting structure with ferromagnetic graphene and confirms that, in this respect, the effect of the sublattice pseudospin degree of freedom in gapped graphene is as important as the spin in a ferromagnetic graphene.

\begin{acknowledgements}
We gratefully acknowledge support by the Institute for Advanced Studies in Basic Sciences (IASBS) Research Council under grants No. G2009IASBS110 and No. G2010IASBS110. We thank A. G. Moghaddam for fruitful discussions. L. M. acknowledges the financial support of Marco Polini and the organizers
of the school NSPM2011 held in Erice, Italy.
\end{acknowledgements}

\end{document}